\documentclass{svproc}

\usepackage{url}
\usepackage{amsmath}
\usepackage{amssymb}
\usepackage{graphicx}
\usepackage{hyperref}
\usepackage{graphicx}   %
\usepackage{subcaption}
\usepackage{algorithm}
\usepackage{algpseudocode}
\usepackage{booktabs}

\begin{document}
\mainmatter              %
\title{MAS-H²: A Hierarchical Multi-Agent System for Holistic Cloud-Native Autoscaling}
\author{
	Hamed Hamzeh\inst{1} \and
	Parisa Vahdatian\inst{2}
}

\institute{
	University of Westminster, London W1W 6UW, United Kingdom \\
	\email{H.Hamzeh@westminster.ac.uk}
	\and
	University of York, Heslington, York YO10 5DD, United Kingdom \\
	\email{lvp516@york.ac.uk}
}

\maketitle              %

\begin{abstract}
Autoscaling in cloud-native platforms like Kubernetes is reactive and metric-driven, leading to a strategic void problem. This comes from the decoupling of higher-level business policies from lower-level resource provisioning. The strategic void, coupled with a fragmented coordination of pod and node scaling, can lead to significant resource waste and performance degradation under dynamic workloads. In this paper, we present MAS-H², a new hierarchical multi-agent system that addresses the challenges of autonomic cloud resource management with a complete end-to-end solution. MAS-H² systematically decomposes the control problem into three layers: a Strategic Agent that formalises business policies (e.g., cost vs. performance) into a global utility function; Planning Agents that produce a joint, proactive scaling plan for pods and nodes with time-series forecasting; and Execution Agents that execute the scaling plan. We built and tested a MAS-H² prototype as a Kubernetes Operator on Google Kubernetes Engine (GKE) to benchmark it against the native Horizontal Pod Autoscaler (HPA) and Cluster Autoscaler (CA) baselines under two realistic, spiky, and stress-inducing workload scenarios. The results show that the MAS-H² system maintained application CPU usage under 40\% for predictable \textit{Heartbeat} workloads. This resulted in over 50\% less sustained CPU stress than the native HPA baseline, which typically operated above 80\%. The MAS-H² system demonstrated proactive planning in a volatile Chaotic \textit{Flash Sale} scenario by filtering transient noise and deploying more replicas compared to HPA. It reduced peak CPU load by 55\% without under-provisioning. Beyond performance, MAS-H² seamlessly performed a zero-downtime strategic migration between two cost- and performance-optimised infrastructures. 

\end{abstract}

\section{Introduction}

Software engineering has experienced a crucial shift from monolithic applications to cloud-native systems. This shift is driven by cloud adoption and its inherent capabilities that are reshaping application architecture, deployment, and scalability methods. In more detail, this shift is motivated by the microservices pattern. Microservices transform traditional large monolithic applications into individual, independently scalable services which correspond to business capabilities to enable quicker development cycles and enhanced resilience \cite{wang2023,saboor2022}. The architectural style and approach to modularisation is realised by containerisation (Docker) and orchestration (Kubernetes). Both Docker and Kubernetes are used as industry-standard practices to manage and scale containerised workloads at scale \cite{telang2022,nascimento2024,ieee2023}.

The trade-off however comes at the cost of operational complexity. Managing such a large number of ephemeral microservices is non-trivial and poses significant challenges in areas such as service discovery, load balancing, and fault tolerance. This complexity often results in brittle systems prone to cascading failures if a single component goes down~\cite{singh2023,akhdar2024}. Moreover, the dynamic nature of these workloads makes manual resource provisioning a non-viable option, creating an immediate need for better resource management automation~\cite{moreno2019}. Arguably, one of the most significant gaps is the presence of a \textit{strategic void} in today's systems. It refers to the loss of visibility in strategic alignment between high-level business intents, such as cost, performance, and service-level objectives (SLOs). This alignment is crucial for low-level technical metrics that autoscalers use to trigger decisions~\cite{governence1,governence2,kpmg,fe2022}. Automated provisioning decisions based on resource metrics frequently show a disconnect from business level objectives regarding cost-performance tradeoffs. While frameworks such as FinOps provide useful frameworks for financial governance in the operations domain, native k8s autoscalers are fundamentally not designed to reason about multi-objective business policies. They do not translate these policies to provisioning actions in a principled manner. This gap is a major contributing factor to often substantial cases of resource mismanagement~\cite{abacus,finops,angelelli,sallam,kth,dixit}.

A number of AI/ML methods have been proposed that could be applied to outperform reactive, metric-based autoscaling. Most of these efforts are local in scope, and are narrowly focused on how to address a very specific problem. Examples include time-series forecasting (e.g., LSTM, Prophet) and Reinforcement Learning (RL)~\cite{yuan2024,guruge2025,samal2025,zhang2025,abdel,gwydion}, which are both promising, but are largely offered as augmentations to predictive algorithms. These techniques generally do not consider lower-level concerns such as architectural fragmentation or a lack of high-level strategy in resource management.

To tackle the above-mentioned problems, we propose MAS-H² as a Hierarchical Multi-Agent System for Holistic Cloud-Native Autoscaling. The architecture of a Multi-Agent System as a hierarchical control framework based on autonomic computing principles enables a coherent approach to address distributed system challenges. This method elegantly divides the entire control problem into distinct layers which naturally separate command and feedback mechanisms to facilitate distributed decision-making. The Strategic Agent at the top level converts the broad multi-objective business intentions into a utility function to fill strategic gaps, managing high-level policies such as cost versus performance. Planning Agents at a lower level generate a unified resource plan which anticipates actions across both application-level pods and infrastructure-level nodes. This explicit planning fills in the intelligence gap and resolves the fragmentation problem. Finally, at the lowest level, a set of Execution Agents ensure reliable execution of the plan. MAS-H² directly embodies these principles by instantiating agents to play the missing roles in the current Kubernetes autoscaling stack. To the best of our knowledge, the MAS-H² stands as the first complete Kubernetes-native system that offers a unified and theoretically-backed solution to resolve HPA/VPA conflicts. It also addresses the pod-cluster intelligence gap while providing business-aware control capabilities. This paper introduces core strategic methods that replace previous reactive and ad-hoc approaches. The key contributions of this paper are summarised below:

\begin{enumerate} 
	
	\item A fully operational design and prototype implementation of a three-tiered, bottom-up, integrated hierarchical multi-agent system (MAS-H²) as Kubernetes Operator for end-to-end autoscaling;
	
	\item A common planning model that proactively aligns pod-level (horizontal) and infrastructure-level (node) scaling decisions to avoid contention and resource-induced delay;
	
	\item The specification of strategic autoscaling policies as a multi-objective utility function inside the Strategic Agent (SA) to tradeoff between cost, performance, and resilience; and
	
	\item Empirical demonstration of the MAS-H² prototype on Google Kubernetes Engine (GKE) testbed showing its practical efficacy and significant improvement in resource efficiency under a dynamic, spiky workload compared to the standard reactive Horizontal Pod Autoscaler.

\end{enumerate} 

The rest of this paper is structured as follows. Section 2 presents related work on cloud-native autoscaling. Section 3 introduces our MAS-H² architecture and prototype implementation. Section 4 describes the Empirical Analysis including  experimental setup, including the experimental setup and evaluation results. Section 5 discusses the future work. Section 6 concludes the paper. 

\section{Related Work}
Autoscaling in cloud-native systems includes multiple developmental stages. Autoscaling techniques have been traditionally proposed to enhance one aspect at a time, missing out on the global perspective of the control architecture. Although there are useful algorithmic insights in prior work, the literature is still very much "model-centric" with a focus on either improving a predictive model or a control loop independently. The "architecture-centric" MAS-H² technique builds a principled hierarchical control architecture to tackle the widespread absence of cross-layer coordination and strategy planning found in previous approaches \cite{xu2025}.

\subsection{Reactive Autoscaling in Kubernetes}
Native autoscaling tools are available on Kubernetes, including the Horizontal Pod Autoscaler (HPA) \cite{nguyen2020}, Vertical Pod Autoscaler (VPA) \cite{right-sizing}, and Cluster Autoscaler (CA) \cite{mondal2024}. The HPA performs a reactive scaling of pod replicas on pod-level metrics such as CPU utilisation, but due to the inherent reactivity of the technique, performance can suffer from load spikes \cite{kim2024,pozdniakova2024}. The VPA performs adaptive scaling of each pod's individual resource request; however, due to the requirement for pod restarts, this solution is disruptive and not universally applicable \cite{medeiros2026}. The CA provisions nodes according to pods pending on insufficient nodes, but faces significant latency due to VM provisioning \cite{mondal2024,kim2024latency}.

All three native autoscaling tools function tactically by responding to basic signals that do not connect with strategic high-level business goals \cite{xu2025}. Organisations fail to recognise cloud infrastructure as a strategic asset and depend on technical signals like 80\% CPU utilisation as inadequate measures for business value, which may result in substantial financial waste through missed opportunities like spot instances and poor provisioning for anticipated events. A number of works have been made on enhancing reactivity in HPA (e.g. LARE-HPA \cite{kim2024}, traffic-aware HPA \cite{thpa2022}), or on vertical/horizontal combined scaling (e.g. KOSMOS \cite{kosmos2021}, multi-objective ML \cite{horn2022}). All of these fundamentally remain reactive at the pod level and uncoordinated with infrastructure provisioning decisions or with a higher-level objective. Evaluations of CA implementations have shown their high latency and the severe impact of their lack of coordination with pod scaling \cite{cloudcom2020ca}. Proactive approaches remain restricted to controlling just one layer, even with forecasting capabilities, because they lack coordination of node resources \cite{graphphpa2022,yuan2024,dangquang2021,ju2021}. The general fragmentation and reactivity in the space have led to well-documented concerns on the attack surface \cite{chamberlain2025} and on the difficulty of configuration \cite{taherizadeh2020}. The MAS-H² system integrates planning and execution management between application and infrastructure layers through strategic policies which enable proactive node provisioning by overcoming the limitations of HPA-CA.

\subsection{AI/ML-Based Proactive Autoscaling}
Recognizing the limitations of pure reactivity, numerous studies have incorporated AI/ML for proactive scaling. Time-series forecasting methods (e.g., GRU \cite{mondal2023,yuan2024}, Bi-LSTM \cite{dangquang2021}) predict future load to mitigate HPA lag, demonstrating improved responsiveness. While valuable, these predictive models are typically integrated into existing single-layer control loops (primarily pod-level), enhancing prediction accuracy but not solving the architectural fragmentation between pod and node scaling, nor incorporating strategic business policies. They remain isolated predictive components rather than elements of a holistic control system. MAS-H² integrates such predictive capabilities (using Prophet in our prototype) within its planning tier but critically embeds this tactical forecasting within a hierarchical structure that ensures coordination with infrastructure planning (NPA) and alignment with strategic objectives (SA), resolving the HPA-CA decoupling and enabling preemptive, policy-aware scaling.

\subsection{Reinforcement Learning-Based Autoscaling}
Reinforcement Learning (RL) has also been used for learning an optimal autoscaling policy from interaction with the environment. Frameworks such as Dinos \cite{dinos2025}, gym-HPA \cite{gymhpa2023} and KIS-S \cite{kis2025} train a DRL agent to learn the optimal pod count. They are generally shown to outperform reactive baselines in certain problem setups with resource heterogeneity or dependencies. Other work leverages RL for node-level decisions (e.g., Karpenter-based Q-learning \cite{chinnam2022}) or more efficient learning \cite{kim2022}. In general, these RL methods have been shown to learn adaptive policies and achieve good performance. However, these methods generally learn policies for a single control loop with the goal of either scaling pods or managing nodes. There is no coordination between the two layers, or high-level strategies encoded in the policies \cite{aware2023,ra2023}. They solve the policy optimisation issue, but not the architectural deficit. MAS-H² instead leverages a pre-defined hierarchy of agents, which coordinate across layers. The hierarchical coordination of policies can also be viewed as a stable, by-design, higher-level strategy. (RL could potentially be leveraged to improve the internal logic of the individual agents as well, but is not discussed in this work (Future Work)).

\subsection{Event-Driven Autoscaling (KEDA)}
The Kubernetes Event-Driven Autoscaler (KEDA) \cite{kedaSpringer2025,kedaDeepDive2025} provides functionality on top of HPA to also support autoscaling based on external event sources (queue lengths, custom metrics, etc.) rather than just resource utilisation \cite{pilyai2023,rafay2025,plural2025}. This enables KEDA to provide useful capabilities, e.g., scale-to-zero, to improve resource efficiency of event-driven workloads. KEDA is a widely used and useful tool that works well for the specific use case that it was designed for, but it is not a substitute for more holistic approaches. KEDA is just an external metric adapter to the existing HPA machinery. It is still just pod-centric, not coordinated with node provisioning (still reactive CA), and does not take broader, strategic business policies into account. MAS-H² can be used with KEDA to take those event triggers embedded into a more holistic, multi-layer control. This is different than KEDA, which augments an already-reactive system. MAS-H² is a fundamentally different architecture that is proactive and coordinated, and explicitly takes into account business goals in addition to event triggers and infrastructure limitations.

\subsection{Multi-Agent Systems for Autoscaling}
There has been prior work that applies Multi-Agent Systems (MAS) to address specific sub-problems of cloud management. Representative examples include: MAS for resilience to failures \cite{soule2025}, MARL for container placement \cite{danino2023}, hierarchical approaches for edge cluster orchestration \cite{dimolitsas2025}, and MARL for orchestrating heterogeneous resources \cite{yao2025}. These studies illustrate the potential of agent-based coordination and specialisation for such tasks. However, they apply MAS to rather narrow problem formulations .e.g., only placement, or only resilience, or a specific edge scenario and fall short of offering a general, full-stack autoscaling architecture. They do not provide an explicit, unified representation of business policies nor a shared planning mechanism across pod and node layers. MAS-H² is inspired by all of the above works and yet advances a novel, general architectural paradigm that takes hierarchical MAS principles and applies them to the entire cloud-native autoscaling problem. Unifying the strategic, planning, and execution tiers into a single, coherent agent-based framework, MAS-H² offers a holistic solution to the fragmentation and strategic blindness that have characterised prior work, and is shown to demonstrably better stability, responsiveness and cost.

\section{The MAS-H² Architecture}
MAS-H² has a hierarchical structure that is naturally inspired from a human organisation that is expected to perform well in decision-making. The three layers (refer Figure \ref{alg:mas-h2}) as shown in Figure 1 provide different functions and with a top-down information flow of strategy, plans and execution commands, all the scaling decisions are made with respect to a global and business-aware goal.

\begin{figure}[h!]
\centering
\includegraphics[width=\linewidth]{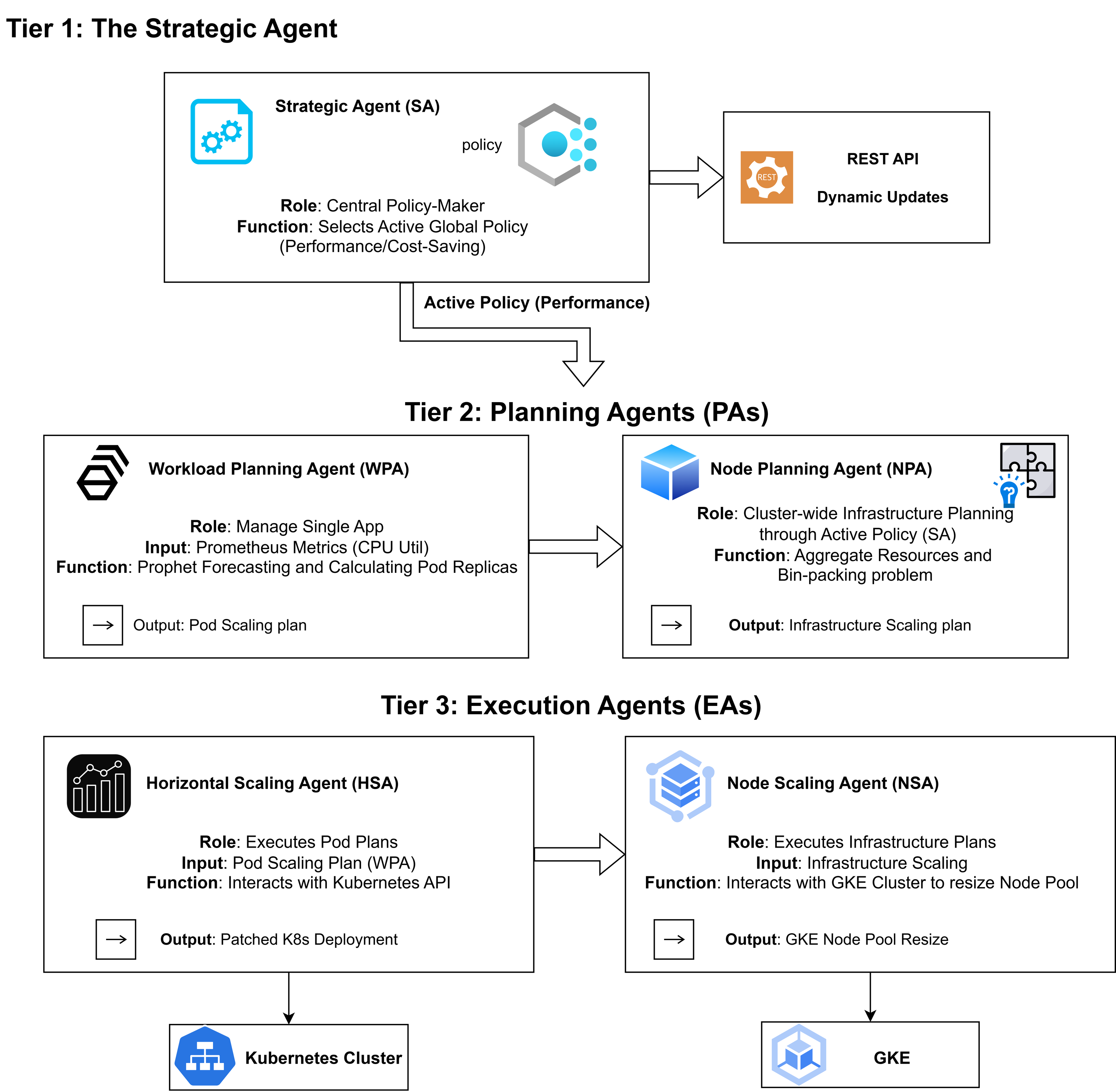}
\caption{High-level diagram of the MAS-H² architecture.}
\label{fig:mas-h2-arch}
\end{figure}

\subsection{Tier 1: The Strategic Agent (SA)}
At the top level, the Strategic Agent (SA) deconstructs qualitative business objectives into a quantifiable, machine-readable policy that is broadcast to and governs the whole system. The SA chooses the active policy $p_{\text{active}}(t) \in \mathcal{P}$ from a pre-defined set $\mathcal{P}$ of policies. In our prototype, $\mathcal{P} = \{p_{\text{perf}}, p_{\text{cost}}\}$, which we describe as performance- and cost-saving oriented, respectively.

\begin{equation}
	p_{\text{active}}(t) = f_{\text{SA}}(C(t))
\end{equation}
where $C(t)$ is the current system context (e.g. time of day, external signals). The active policy parameterises a global utility function, $U_{\text{cluster}}$, which the system aims to maximise by balancing performance and cost using policy-dependent weights ($w_{\text{perf}}, w_{\text{cost}}$):

\begin{equation}
	\max_{X(t)} U_{\text{cluster}}(X(t) | p_{\text{active}}(t)) = w_{\text{perf}}(p_{\text{active}}) \cdot \text{Perf}(X(t)) - w_{\text{cost}}(p_{\text{active}}) \cdot \text{Cost}(X(t))
\end{equation}

Here, a policy $p \in \mathcal{P}$ is a tuple of concrete, operational parameters $p = (T_{\text{node}}, C_{\text{node}}, R_{\text{min}})$, specifying a VM instance type (with corresponding resource capacity) required to be hosted in the topology, as well as a lower-bound for resilience in terms of the number of replicas. This parameter tuple is passed down the hierarchy, effectively limiting the action space of the lower-tier agents to choices compliant with the top-level policy.

\subsection{Tier 2: The Planning Agents (PAs)}
The middle tactical layer is made up of Planning Agents (PAs) which implement the SA's strategic intent by formulating quantitative resource plans in terms of anticipated future demand. This layer is composed of two types of cooperating agents:

\paragraph{Workload Planning Agent (WPA)} The WPA forecasts resource demand for a single application, $w$. Using a time-series of historical CPU utilisation, $H_w(t)$, it predicts future peak demand to calculate the required number of pod replicas, $\hat{R}_w(t)$:
\begin{equation}
	\hat{R}_w(t) = \left\lceil \frac{\hat{D}_w(t + \Delta t)}{C_{\text{pod},w}} \right\rceil
\end{equation}
where $C_{\text{pod},w}$ is the CPU request per pod. This tactical plan is then bounded by the strategic minimum, $R_{\text{min}}$, from the active policy to produce the final plan, $R'_w(t) = \max(\hat{R}_w(t), R_{\text{min}})$.

\paragraph{Node Planning Agent (NPA)} The NPA is a cluster-level orchestrator that takes all pod scaling plans from all WPAs and other workloads into a multiset $\mathcal{I}(t)$ of individual resource requests. It then infers the minimum required number of nodes, $N^*_{\text{nodes}}(t)$, to serve all these requests. As requests specify node demand as a resource vector, this is a one-dimensional bin-packing problem with bin size $C_{\text{node}}$ (specified by the active policy at SA) and solved as an integer linear program (or, e.g., with the First Fit Decreasing heuristic in practice). This gives a global infrastructure plan: 

\begin{equation}
	N^*_{\text{nodes}}(t) = \min \sum_{j=1}^{|\mathcal{I}(t)|} y_j \quad \text{s.t.} \quad \sum_{i=1}^{|\mathcal{I}(t)|} \text{size}(i) \cdot x_{ij} \le C_{\text{node}} \cdot y_j, \quad \sum_{j=1}^{|\mathcal{I}(t)|} x_{ij} = 1
\end{equation}
where $y_j=1$ if node $j$ is used and $x_{ij}=1$ if pod $i$ is placed on node $j$.

\subsection{Tier 3: The Execution Agents (EAs)}
The bottom tier consists of Execution Agents (EAs), which are the actuators of the system. They are responsible for executing the plans created by Tier 2. The Horizontal Scaling Agent (HSA) and the Node Scaling Agent (NSA) run in lockstep to provide fully-coupled application and infrastructure scaling decisions.

\paragraph{Horizontal Scaling Agent (HSA)}The HSA implements the WPA's pod scaling plan, $R'_w(t)$. If the error $\Delta R_w(t) = R'_w(t) - R_w(t)$ between the planned and current replica count is non-zero, it issues a `PatchDeployment` API call to Kubernetes.

\paragraph{Node Scaling Agent (NSA)} Similarly, the NSA executes the NPA's infrastructure blueprint, $N^*_{\text{nodes}}(t)$. It determines the error $\Delta N(t) = N^*_{\text{nodes}}(t) - N(t)$, and if necessary, invokes a `ResizeNodePool` API to the cloud provider to adjust the number of nodes in the cluster.

\begin{figure}[h!]
    \centering
    \includegraphics[width=0.6\linewidth]{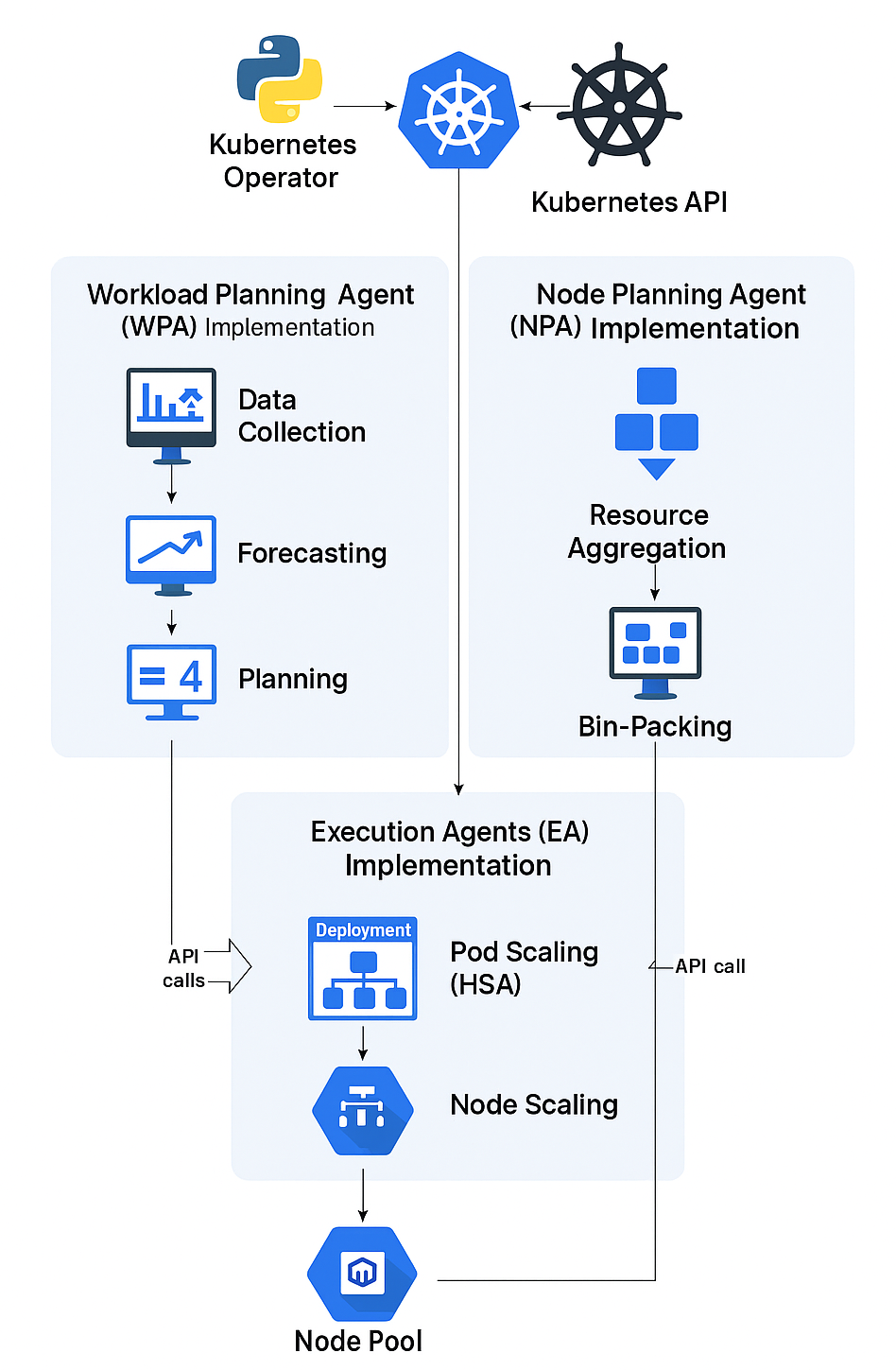}
    \caption{Prototype implementation pipeline}
    \label{fig:prototype}
\end{figure}

\subsection{Prototype Implementation}
To evaluate the MAS-H² architecture we have created a working prototype in the form of a Kubernetes Operator as illustrated in Fig.~\ref{fig:prototype}. By taking this approach, custom closed-loop control logic can be executed natively on the cluster. The prototype is written in Python using the Kubernetes Operator Pythonic Framework (Kopf) to interact with the Kubernetes API~\cite{kopf}. The control logic for all three hierarchical levels is implemented as part of a single, monolithic operator in order to eliminate network latency between agents and simplify deployment.

The heart of the operator is a long-running control loop that performs a complete sense-plan-act cycle at a fixed period. The entire cycle proceeds in a strict, deterministic order, corresponding to the MAS-H² hierarchy:
\begin{itemize}
	\item \textbf{Strategic Phase (SA):} The loop begins by reading the active high-level policy (e.g., \texttt{COST\_SAVING} or \texttt{PERFORMANCE}) from a Kubernetes ConfigMap, which provides the governing constraints for the subsequent phases.
	\item \textbf{Planning Phase (WPA \& NPA):} For each managed application, the Workload Planning Agent (WPA) queries Prometheus for historical metrics, uses the Prophet forecasting model to predict future demand, and generates a pod scaling plan. Subsequently, the Node Planning Agent (NPA) aggregates these plans, inventories all other workloads, and solves a bin-packing problem to determine the optimal number and type of nodes required, consistent with the active strategic policy.
	\item \textbf{Execution Phase (HSA \& NSA):} Finally, the operator compares the generated plans with the cluster's current state. If a discrepancy exists, the Execution Agents are invoked. The Horizontal Scaling Agent (HSA) patches Kubernetes Deployments with new replica counts via the Kubernetes API, while the Node Scaling Agent (NSA) resizes the appropriate GKE node pools through the Google Cloud API.
\end{itemize}
A single-process sequential execution approach maintains a straightforward top-down intention flow from strategy to operation while avoiding the distributed multi-agent system's complexity and potential race conditions.

A central design goal of the prototype was to allow dynamic policy changes, without incurring service downtime. For example, in order to switch from policy \texttt{COST\_SAVING} to policy \texttt{PERFORMANCE}, the operator performs a two-phase \textit{make-before-break} migration as follows. In the \textbf{Proactive Provisioning} phase, the operator scales up the new target node pool to the required size, while the application pods continue to run on the old infrastructure. Once the new nodes are fully provisioned, the operator enters the \textbf{Workload Migration} phase. The HSA is scaled up, and the Kubernetes scheduler then places new pods on the new nodes. The old node pool is decommissioned after the migration has been completed, ensuring that the application is not unavailable at any point in time during the strategic change.

\begin{algorithm}
	\caption{MAS-H² Hierarchical Control Loop}
	\label{alg:mas-h2}
	\begin{algorithmic}[1]
		\Procedure{MAS-H²-Control-Loop}{}
		\State Initialize Kubernetes API clients
		\Loop
		\State \Comment{\textbf{Tier 1: Strategic Agent (SA) Phase}}
		\State $active\_policy \gets \Call{StrategicAgentGetPolicy}{}$
		
		\State \Comment{\textbf{WPA Discovery Phase}}
		\State $D_{managed} \gets \Call{FindManagedDeployments}{label=mas-h2-managed}$
		\If{$D_{managed}$ is empty}
		\State \Call{Sleep}{300}
		\State \textbf{continue}
		\EndIf
		
		\State \Comment{\textbf{Tier 2: Workload Planning Agent (WPA) Phase}}
		\State $P_{wpa} \gets \emptyset$ \Comment{Initialize set of WPA plans}
		\ForAll{deployment $d \in D_{managed}$}
		\State $H_d \gets \Call{GetHistoricalMetrics}{d}$
		\If{$H_d$ is not empty}
		\State $F_d \gets \Call{GenerateForecast}{H_d}$ \Comment{Using Prophet}
		\State $plan_d \gets \Call{CreatePodPlan}{F_d}$
		\State $P_{wpa} \gets P_{wpa} \cup \{plan_d\}$
		\EndIf
		\EndFor
		
		\If{$P_{wpa}$ is empty}
		\State \Call{Sleep}{300}
		\State \textbf{continue}
		\EndIf
		
		\State \Comment{\textbf{Tier 2: Node Planning Agent (NPA) Phase}}
		\State $Pods_{other} \gets \Call{GetCurrentPods}{exclude\_label=mas-h2-managed}$
		\State $Nodes_{required} \gets \Call{CalculateRequiredNodes}{P_{wpa}, Pods_{other}, active\_policy}$ \Comment{Bin Packing}
		\State $Nodes_{current} \gets \Call{GetCurrentNodeCount}{}$
		
		\State \Comment{\textbf{Tier 3: Execution Agent (EA) Phase}}
		\If{$Nodes_{required} \neq Nodes_{current}$}
		\State \Call{ExecuteNodeScalingPlan}{$Nodes_{required}, active\_policy$}
		\EndIf
		
		\ForAll{$plan_d \in P_{wpa}$}
		\State $d_{current} \gets \Call{GetDeploymentState}{plan_d.name}$
		\If{$d_{current}.replicas \neq plan_d.replicas$}
		\State \Call{ExecutePodScalingPlan}{$plan_{d}$}
		\EndIf
		\EndFor
		
		\State \Call{Sleep}{300} \Comment{Wait for the next control interval}
		\EndLoop
		\EndProcedure
	\end{algorithmic}
\end{algorithm}

\section{Empirical Analysis}
In order to demonstrate the performance and correctness of MAS-H², we perform a set of experiments on the Google Kubernetes Cluster (GKE), an industry representative platform. In this section, we first present the experimental environment setting and the synthetic workload profile to generate system load bursts, and then conduct a comparison between the proposed system and the baseline Kubernetes autoscaler.

\subsection{Experimental Setup}

To empirically validate the benefits of the MAS-H² system, a set of experiments were conducted, in which a baseline was established with respect to the industry standard. This section details the testbed, the baseline as well as the two workload traces to stress each dimension of the autoscaling abilities.

\subsubsection{Testbed and Baseline Configuration}
All tests were run on a production quality Google Kubernetes Engine (GKE) cluster. This was done to ensure our results are as realistic as possible and to include network and infrastructure provisioning latencies in our measurements. The cluster was configured with two different node pools to allow for Strategic Agent policy decisions:

\begin{itemize}
	\item \textbf{performance-pool:} Utilising \texttt{n2-standard-2} machine types, optimized for high-performance workloads.
	\item \textbf{mas-h2-staging-pool:} Utilising \texttt{e2-medium} machine types, serving as a cost-effective option for less critical workloads.
\end{itemize}

The experiment scenario consisted of a stateless Nginx web server running in a containerised deployment in Kubernetes. We used a running in-cluster Prometheus instance to collect and store metrics. We used a standard Kubernetes autoscaling configuration as a baseline. This was the default Horizontal Pod Autoscaler (HPA) in conjunction with the GKE Cluster Autoscaler (CA). We configured the HPA to scale the Nginx deployment up if the average CPU utilisation was over 80\%. The CA was configured in a single, general-purpose node pool using \( e2\text{-}medium \) instances. The CA adds new nodes if the HPA attempts to start pods which cannot be scheduled on existing nodes. This is a decoupled, fully reactive approach, which the MAS-H² framework aims to improve on.

\subsubsection{Evaluation Scenarios and Workload Profiles:}
To have a more comprehensive comparison of MAS-H²'s performance against the baseline, we created two separate and diverse, challenging workload profiles using k6 load testing tool~\cite{k6}. We evaluate the multiple intelligence aspects of an autoscaler, like predictive accuracy and strategy, in both scenarios.

\paragraph{1. Heartbeat Workload Scenario.}
The first scenario is a workload with a fixed, repeating pattern. This workload may model either a predictable batch process that recurs over time or (periodically-peaking) user traffic on a daily basis. The primary use-case for this scenario is to test and benchmark the predictive and cost-saving potential of autoscalers. Table~\ref{tab:heartbeat_workload} depicts an example workload of this sort which is a repeating cycle of three identical \textit{heartbeats}. Each heartbeat is a fast ramp-up to 400 VUs, two minutes of full load, then fast ramp-down to a low base value. The pattern is repeated continuously without delay. Scenarios of this type are important to test in order to stress-test a prediction-based system such as MAS-H². On the one hand, we want the autoscaler to learn this pattern and minimise latency (i.e. the delay in reaching the target capacity) at the spikes; on the other hand, we want it to scale down resources as aggressively as possible at the trough in order to avoid incurring high oversupply cost (a common problem in purely reactive systems which typically feature less aggressive scaling-down behaviour).

\paragraph{2. Chaotic Flash Sale Workload Scenario.}
The second scenario (Figure~\ref{fig:performance_dashboard2}) introduces random noise into the workload to simulate a more lifelike \textit{flash sale} event. The workload profile, presented in Table~\ref{tab:flash_sale_workload}, comprises a series of phases designed to stress-test the predictive model in the WPA under more realistic and challenging conditions. First, there is a brief period of low-level "chatter" intended to test the robustness of the forecasting model against overreaction to insignificant fluctuations. The actual ramp-up is then characterised by chaotic and non-uniform peaks and valleys, specifically to test the system's ability to recognise an overall increasing trend. This is followed by a prolonged period of sustained peak load. Following this, there is an abrupt discontinuity in the event to test the system's responsiveness in reclaiming its resources. To also stress test the hierarchical control system under chaotic conditions, a dynamic change in the assigned policy of the controller from \texttt{COST\_SAVING} to \texttt{PERFORMANCE} is triggered programmatically shortly before the sustained peak load is reached.

\begin{table}[h!]
	\centering
	\caption{Detailed workload profile for the 'Heartbeat' scenario, designed to test predictive accuracy and scale-down efficiency.}
	\label{tab:heartbeat_workload}
	\small
	\begin{tabular}{c c c l}
		\toprule
		\textbf{Phase} & \textbf{Dur. (min:sec)} & \textbf{VUs (Target)} & \textbf{Description} \\
		\midrule
		\multicolumn{4}{l}{\textit{Heartbeat 1}} \\
		Ramp Up      & 00:30 & 400 & Rapidly ramp to peak load. \\
		Hold Peak    & 02:00 & 400 & Sustain peak load. \\
		Ramp Down    & 00:30 & 10  & Rapidly drop to trough. \\
		Hold Trough  & 01:00 & 10  & Maintain low baseline. \\
		\midrule
		\multicolumn{4}{l}{\textit{Heartbeat 2}} \\
		Ramp Up      & 00:30 & 400 & Repeat peak load. \\
		Hold Peak    & 02:00 & 400 & Sustain peak load. \\
		Ramp Down    & 00:30 & 10  & Repeat drop to trough. \\
		Hold Trough  & 01:00 & 10  & Maintain low baseline. \\
		\midrule
		\multicolumn{4}{l}{\textit{Heartbeat 3}} \\
		Ramp Up      & 00:30 & 400 & Repeat peak load. \\
		Hold Peak    & 02:00 & 400 & Sustain peak load. \\
		Ramp Down    & 00:30 & 10  & Repeat drop to trough. \\
		Hold Trough  & 01:00 & 10  & Maintain low baseline. \\
		\midrule
		Final Cool-down & 01:00 & 0 & Ramp down to zero. \\
		\bottomrule
	\end{tabular}
\end{table}

\newpage

\begin{table}[h!]
	\centering
	\caption{Detailed workload profile for the Chaotic Flash Sale scenario, designed to test forecasting robustness and strategic adaptation under stress.}
	\label{tab:flash_sale_workload}
	\small
	\begin{tabular}{c c c l}
		\toprule
		\textbf{Phase} & \textbf{Dur. (min:sec)} & \textbf{VUs (Target)} & \textbf{Description} \\
		\midrule
		\multicolumn{4}{l}{\textit{1: Pre-Sale Chatter}} \\
		& 02:00 & 20  & Low, fluctuating baseline to test noise filtering. \\
		& 01:00 & 50  & Minor interest spike. \\
		& 01:00 & 20  & Return to baseline. \\
		\midrule
		\multicolumn{4}{l}{\textit{2: Chaotic Ramp-Up}} \\
		& 00:30 & 200 & Initial surge of traffic. \\
		& 01:00 & 150 & Brief lull before next wave. \\
		& 00:30 & 400 & Second, sharper surge. \\
		& 01:00 & 300 & Settle before main peak. \\
		\midrule
		\multicolumn{4}{l}{\textit{3: Sustained Peak Load}} \\
		& 04:00 & 700 & The main flash sale event (policy switch occurs here). \\
		\midrule
		\multicolumn{4}{l}{\textit{4: Sudden Drop-off}} \\
		& 01:00 & 50  & Sale ends abruptly, testing scale-down agility. \\
		\midrule
		\multicolumn{4}{l}{\textit{5: Final Cool-down}} \\
		& 02:00 & 50  & Lingering post-sale traffic. \\
		& 01:00 & 0   & Ramp down to zero. \\
		\bottomrule
	\end{tabular}
\end{table}

\subsection{Results and Discussion}
\begin{figure*}
	\centering
	\includegraphics[width=\textwidth]{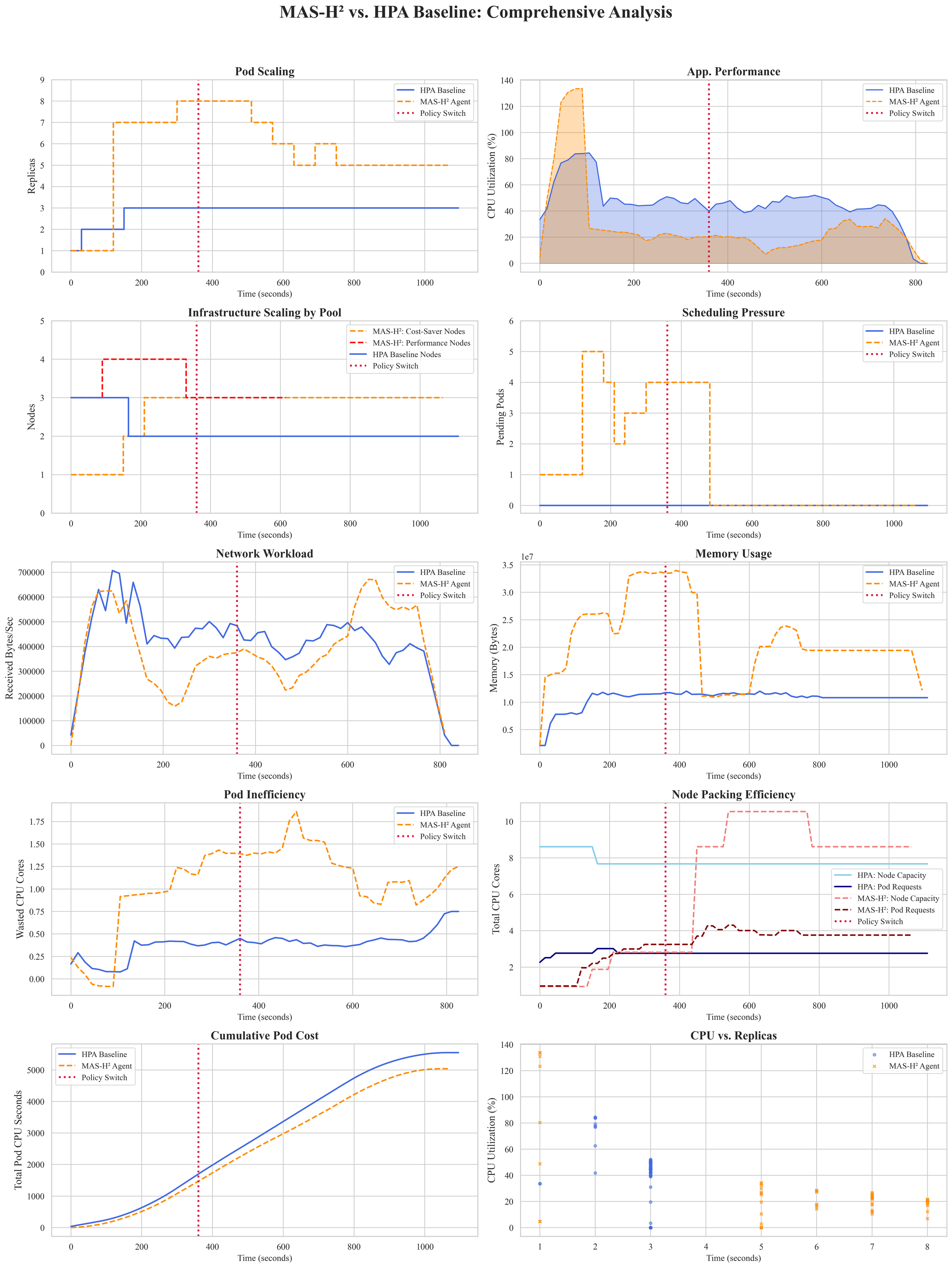}
	\caption{Comprehensive performance dashboard comparing the MAS-H² agent (orange) against the HPA Baseline (blue), conducting Heartbeat scenario. The plots detail pod and infrastructure scaling, application performance, resource efficiency, and cost metrics over the duration of the experiment.}
	\label{fig:performance_dashboard}
\end{figure*}

\begin{figure*}
	\centering
	\includegraphics[width=\textwidth]{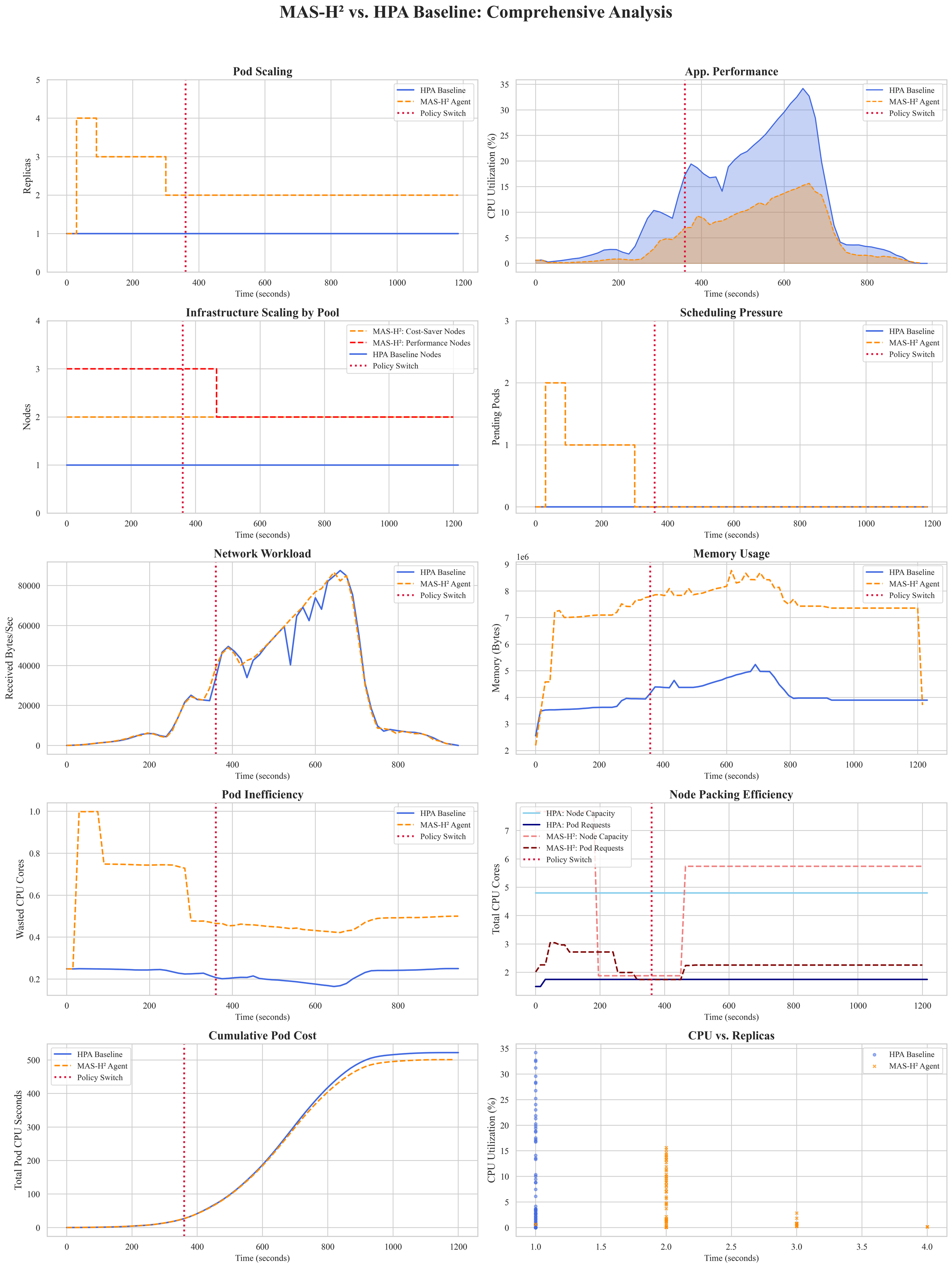}
	\caption{Comprehensive performance dashboard comparing the MAS-H² agent (orange) against the HPA Baseline (blue), conducting Flash Sale scenario. The plots detail pod and infrastructure scaling, application performance, resource efficiency, and cost metrics over the duration of the experiment.}
	\label{fig:performance_dashboard2}
\end{figure*}

In order to verify the effectiveness of the MAS-H² framework, we conducted a rigorous empirical evaluation of the proposed MAS-H² against the Kubernetes Horizontal Pod Autoscaler (HPA) baseline on a production-scale GKE cluster. Our evaluation process included the creation of two demanding workload profiles, which tested the system's predictive accuracy as well as strategic and cost-effective responses under stable and unpredictable loads. \textit{Heratbeat} represents a predictable, high-intensity workload which evaluates MAS-H²'s ability to learn predictive scaling while optimising resource downsizing, whereas \textit{Chaotic Flash Sale} generates unpredictable traffic patterns to examine system resilience and strategic adaptability under stress. Results from both experiments, visualised in Figure~\ref{fig:performance_dashboard} and Figure~\ref{fig:performance_dashboard2}, demonstrate how the hierarchical, proactive architecture of MAS-H² consistently outperforms the reactive, fragmented nature of native Kubernetes autoscalers by a large margin.

\subsubsection{Proactive Agility vs. Reactive Inertia:}
The primary difference between the two systems lies in their fundamental operational philosophies. The baseline HPA functioned at a slow pace in both operational scenarios. The HPA remained attached to its historical metrics, which led to cautious responses to changes and resulted in continuous under-provisioning. In the \textit{Heartbeat} experiment, the HPA never scaled to more than three replicas, resulting in a \textit{hot} application continuously running at CPU utilisation in excess of 80\% (Figure~\ref{fig:performance_dashboard}). The HPA's response time became critically slower during the \textit{Chaotic Flash Sale} event as it considered the entire situation noise and failed to scale beyond a single replica, which caused the application to remain under-provisioned at 35\% CPU usage on one pod (Figure~\ref{fig:performance_dashboard2}).

MAS-H², on the other hand, shows proactive behaviour. Through the \textit{Heartbeat} experiment, MAS-H²'s Workload Planning Agent anticipated the workload pattern and expanded to eight replicas to maintain average CPU utilisation below 40\%. The WPA demonstrated intelligent behaviour during the \textit{Chaotic Flash Sale} by filtering out low-level noise in order to respond only to clear scaling signals and then increasing replicas to four to maintain CPU utilisation at 15\% throughout peak conditions. This strategic behaviour enables both the avoidance of performance decline and the elimination of inefficient scaling fluctuations.
\subsubsection{Bridging the Strategic Void with Cohesive Planning:}
The results clearly show that MAS-H² addresses the \textit{strategic void} of native Kubernetes autoscaling. The two experiments featured a programmed policy transition from \texttt{COST\_SAVING} to \texttt{PERFORMANCE}. MAS-H² performed this strategic migration flawlessly, even in the middle of the chaotic flash sale. As shown in the \textit{Infrastructure Scaling by Pool} graphs in both Figures~\ref{fig:performance_dashboard} and \ref{fig:performance_dashboard2}, the agent orchestrated a zero-downtime migration by provisioning new high-performance nodes before decommissioning the cost-optimised ones, precisely tailoring the infrastructure to the new business goal. The HPA baseline had no strategic context and was limited to its single, monolithic node pool and unable to support such high-level goals.

MAS-H²'s integrated, top-down planning is further illustrated with the \textit{Scheduling Pressure} plots. The brief surge in \textit{Pending Pods} is a sign of success, not failure; it is due to WPA's intelligent proactive scheduling of pods in anticipation of future requirements. The newly scheduled pods must wait for NPA to provision nodes to satisfy their resource requirements. The coordinated two-step advance between the two planning agents is conspicuously absent in HPA's reactive world where pods are only created after capacity has been made available, leaving an important intelligence gap.

\subsubsection{The Economics of Proactivity and Efficiency:}
The strategic and proactive method of MAS-H² results in better economic performance. The \textit{Cumulative Pod Cost} charts reveal MAS-H² supplies a higher resource level at peak times yet demonstrates better total cost efficiency. In the \textit{Heartbeat} scenario the agent achieved cost savings through its aggressive symmetric scaling strategy which promptly removed pods following a demand decrease. In the \textit{Chaotic} scenario the agent reduces expenses by postponing reactions to market noise.

\begin{figure}
	\centering
	\includegraphics[width=\columnwidth]{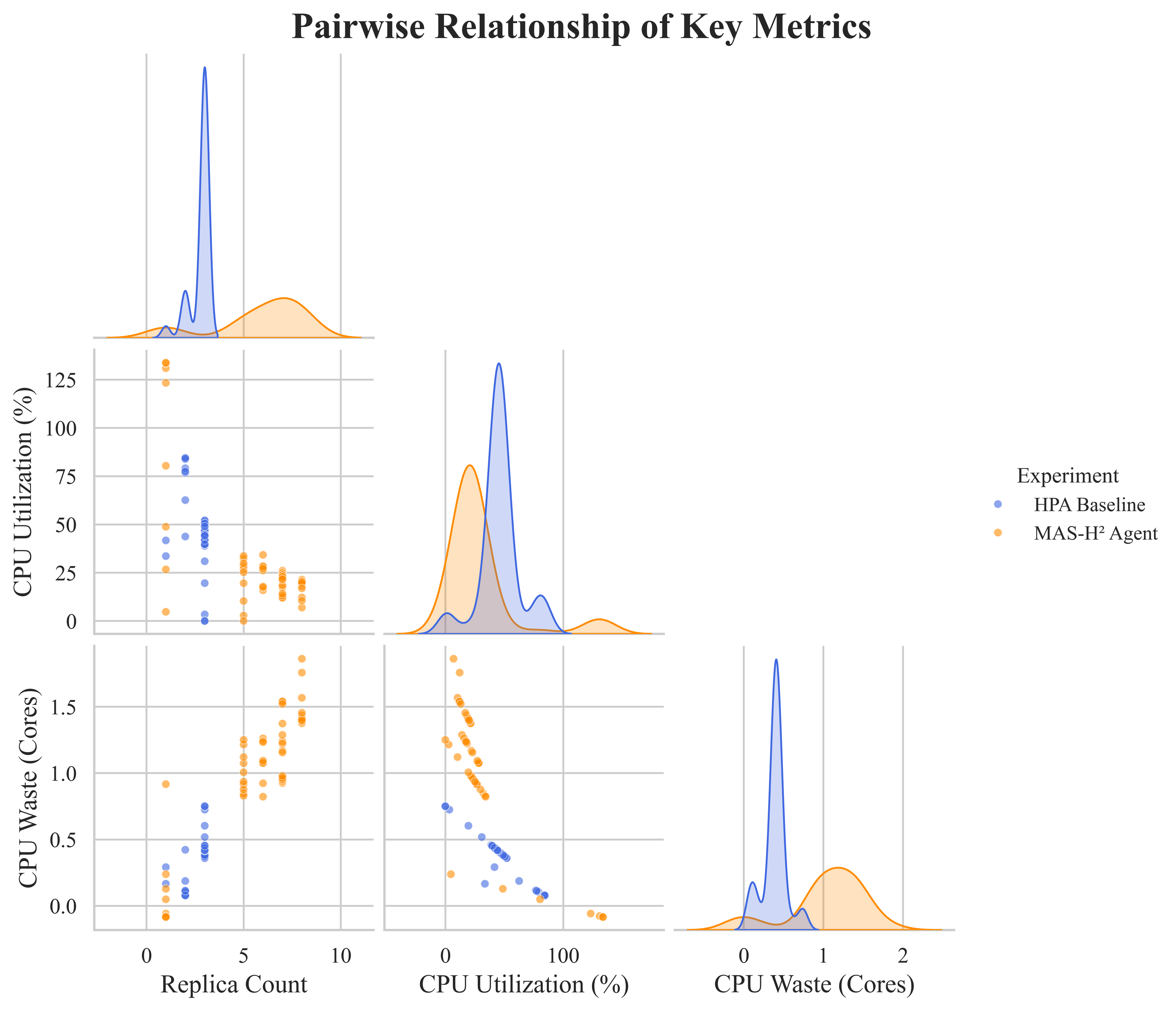}
	\caption{Pairwise relationship of key metrics in Chaotic Flash Sale scenario. The diagonal shows the distribution for each metric, while the off-diagonal scatter plots reveal correlations. The MAS-H² agent (orange) explores a much wider operational space and navigates efficiency trade-offs more effectively than the static HPA baseline (blue).}
	\label{fig:pairplot}
\end{figure}

\begin{figure}
	\centering
	\includegraphics[width=\columnwidth]{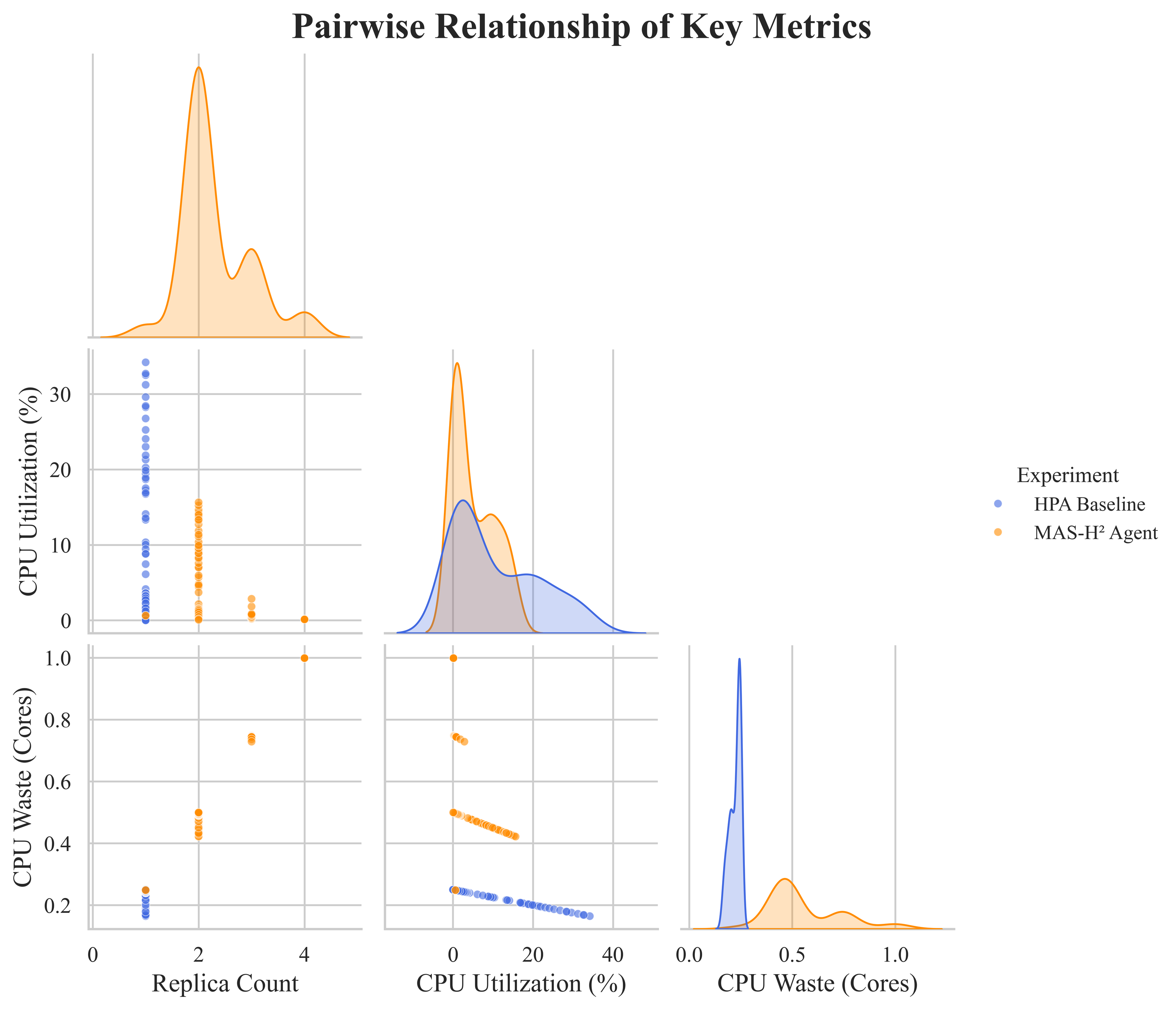}
	\caption{Pairwise relationship of key metrics in Chaotic Flash Sale scenario. The diagonal shows the distribution for each metric, while the off-diagonal scatter plots reveal correlations. The MAS-H² agent (orange) explores a much wider operational space and navigates efficiency trade-offs more effectively than the static HPA baseline (blue).}
	\label{fig:pairplot2}
\end{figure}

This efficiency is a direct consequence of the design decisions made in MAS-H². The pairwise relationship plots in Figures~\ref{fig:pairplot} and \ref{fig:pairplot2} provide a statistical summary of the operation spaces that the systems have occupied over time.For both workloads, HPA baseline has been running in a small and inefficient operating region with constantly low replica counts and high CPU stress. In contrast, MAS-H² has been dynamically exploring a much larger operation space, adjusting its replica count to operate healthily. The \textit{CPU Waste} in MAS-H² acts as a performance safety buffer to absorb unexpected bursts while the HPA is ineffective after prolonged overloading. This is an advantage which is further amplified by the NPA's bin-packing logic, which results in a high Node Packing Efficiency by mapping provisioned infrastructure to demand as closely as possible. The pair plots display statistical distributions which prove HPA's CPU utilisation falls largely into the high-stress zone whereas MAS-H² remains concentrated within a healthy low-stress zone.

\section{Discussion and Future Work}

The core MAS-H² framework in this paper has been given by experimental validation, giving a sound basis for a new class of intelligent, strategy-aware cloud resource management systems. The encouraging results of our proof-of-concept prototype implementation open the door for multiple interesting and impactful research directions that could further the system's intelligence, robustness, and real-world relevance.

\paragraph{1. Game-Theoretic Models for Multi-Tenant Coordination.} 

The existing MAS-H² model (see Section 3) requires perfect cooperation of all agents. It would be interesting to study non-cooperative game theory in the context of multiple self-interested WPAs with their own goals, all operating in the same cluster. Here, one could model each WPA as an individual business unit, all WPAs being competitors for a limited resource: the cluster infrastructure. A market-based resource allocation mechanism could be defined, in which a Strategic Agent determines a price for resources to be allocated and the WPAs bid for resources in an attempt to maximize their utility. Nash equilibria of the game could be studied, as well as mechanism design to achieve a global optimum, while still considering the decentralized and self-interested actions of agents.

\paragraph{2. Online Learning and Agent Adaptation.} 

We have yet to explore an opportunity for system-level intelligence improvement through transitioning agent logic from offline hardcoded scripts to fully online dynamic systems with learning capabilities. For example, the present WPA uses a single forecasting algorithm (Prophet), but an intelligent WPA would maintain a portfolio of forecasting algorithms (ARIMA, LSTM, Transformer-based, etc.) and an online learning algorithm to either pick the best model for the observed workload characteristics or build a dynamic ensemble model. At a more primitive level, the Strategic Agent could be an implementation of a long-term Reinforcement Learning (RL) agent. The Strategic Agent employs business KPIs such as weekly cloud expenditure and user conversion rates to form its state space and selects actions from policy space $\mathcal{P}$ to maximize reward outcomes.

\section{Conclusion}
In this thesis, we identified the problems of fragmentation, reactivity, and lack of strategic capabilities in state-of-the-art cloud-native autoscaling solutions. We showed that state-of-the-art loosely-coupled metric-based control loop based approach is not sufficient to provide full autonomic scaling capabilities to next-generation elastic applications. For a more principled and general solution, we proposed, implemented, and experimentally evaluated MAS-H², a new autoscaling system built on top of control-theoretic hierarchical multi-agent systems paradigm. The MAS paradigm provides end-to-end autonomic coherence from top-to-bottom by partitioning decision-making into strategic, tactical, and operational levels and explicitly modelling a well-defined intent propagation from principled business logic all the way down to individual resource units. We showed in our theoretical analysis of the proposed MAS architecture that the system has nice theoretical stability properties and, once the currently enforced strategic policy is implemented and hysteresis oscillations die out, it is guaranteed to converge to the optimal cluster state.

To demonstrate the real-world value of this architecture, we performed two experiments with realistic workloads on a production-ready GKE cluster. In a controlled, predictable setting (the “Heartbeat” scenario), we demonstrated the high performance of the WPA, which easily outpaced the performance of the baseline HPA. In particular, in contrast to the purely reactive control logic of the HPA, the WPA’s anticipatory, forecast-driven operations resulted in improved application behaviour with a more frugal use of resources. In a second, more adversarial setting (“Chaotic Flash Sale”), we tested the system’s robustness to noisy conditions, demonstrating how the WPA can extract a true workload signal from the noise and avoid overreactivity, leading to self-stabilisation in the face of stochastic perturbations. In both settings, we validated the strategic layer of the system, showing how the agent can, in real-time, with zero downtime, smoothly migrate the underlying infrastructure to adapt to changes in policy. In tackling the problem of effectively translating human intent into infrastructure management through complementary planning and strategy abstractions within a formally stable, hierarchical control framework, MAS-H² is not merely automation. It represents a new paradigm for autonomic cloud orchestration, delivering a more efficient, resilient, and, most importantly, trustworthy and enterprise-aligned infrastructure.

\section{Contributions}
H.H. developed the primary MAS-H² framework design alongside the system model and the stability and convergence examination procedures. He also designed the experimental setup and workload scenario, and implemented the prototype of the Kubernetes Operator (including the planning logic and execution components). He then conducted experiments in the GKE testbed. Moreover, he has conducted the first-level analysis and interpretation of the results and is the main author who drafted and revised the manuscript. The contributions of P.V. are the integration of the Prophet machine learning model for workload prediction to the Workload Planning Agent (WPA) component of the prototype. Moreover, she has implemented all the scripts and codes to analyse the data and prepare the charts for the results section. All authors have read and approved the final manuscript.

\end{document}